\documentclass{article}

\usepackage{arxiv}

\usepackage[utf8]{inputenc} % allow utf-8 input
\usepackage[T1]{fontenc}    % use 8-bit T1 fonts
\usepackage{hyperref}       % hyperlinks
\usepackage{url}            % simple URL typesetting
\usepackage{booktabs}       % professional-quality tables
\usepackage{amsfonts}       % blackboard math symbols
\usepackage{nicefrac}       % compact symbols for 1/2, etc.
\usepackage{microtype}      % microtypography
\usepackage{amsmath}
\usepackage{amssymb}
\usepackage{graphicx}

\usepackage[round,authoryear]{natbib} 

\usepackage{doi}

\title{SarDub19: An Error Estimation and Reconciliation Protocol}

\author{ \href{https://orcid.org/0000-0003-2697-1756}{\includegraphics[scale=0.06]{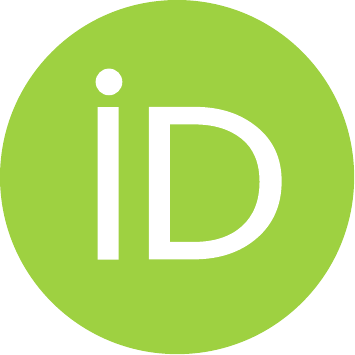}\hspace{1mm}Miralem Mehic} \\
	Department of Telecommunications, \\Faculty of Electrical Engineering, \\ University of Sarajevo, \\Zmaja od Bosne bb, 71000 Sarajevo, \\Bosnia and Herzegovina;\\ \\
    VSB-Technical University of Ostrava, \\17.listopadu 15, \\708 00 Ostrava-Poruba, \\Czech Republic \\
	\texttt{miralem.mehic@ieee.org} \\
	\And
	\href{https://orcid.org/0000-0003-1371-2683}{\includegraphics[scale=0.06]{orcid.pdf}\hspace{1mm}Harun Siljak} \\
	CONNECT Centre, \\Dunlop Oriel House, \\
34 Westland Row, \\
Trinity College Dublin, \\ the
University of Dublin,\\Dublin 2,\\
Ireland
 \\
	\texttt{harun.siljak@tcd.ie} \\
}

\renewcommand{\shorttitle}{SarDub19: An Error Estimation and Reconciliation Protocol}

\hypersetup{
pdftitle={SarDub19: An Error Estimation and Reconciliation Protocol},
pdfsubject={Quantum Cryptography},
pdfauthor={Miralem Mehic and Harun Siljak},
pdfkeywords={Quantum Cryptography, Protocols, Error Estimation, Error Reconciliation, Security},
}

\begin{document}
\maketitle

\begin{abstract}
Aside from significant advancements in the development of optical and quantum  components, the performance of practical quantum key distribution systems is largely determined by the type and settings of the error key reconciliation procedure. It is realized through public channel and it dominates the communication complexity of the quantum key distribution process. The practical utilization significantly depends on the computational capacities that are of great importance in satellite-oriented quantum communications. Here we present SarDub19 error key estimation and reconciliation protocol that improves performances of practical quantum systems.
\end{abstract}

% keywords can be removed

\section{Introduction}

Quantum Key Distribution (QKD) relies on physical laws to distribute cryptographic keys between distant parties in an information-theoretically secure (ITS) way~\cite{bennett1984quantum,brassard1994}. Interest in QKD technology is growing as it increasingly proves its maturity through practical international platforms~\cite{Mehic2020b,SasakiM20111,Peev2009,chen2009,Wang2014,Elliott2005}, commercial products~\cite{Hosseinidehaj2018}, increased rates and distances~\cite{Lucamarini2018a,Wehner2018a,Korzh2015} as well as resistance to hacking attacks~\cite{Shor2000,Mayers2001,Renner2005a,gisin2002quantum}. 

With efforts to overcome the distance constraints using satellite connections, parties need to consider significant losses in optical channel, limited time to establish secret key due to periodic satellite availability where communication performances put additional constraint~\cite{Bedington2017a,Liao2017a,Yin2017,Vallone2014}. 

To provide information-theoretic secrecy (ITS), QKD uses non-deterministic random number generators that rely on the quantum state of matter as a source of entropy~\cite{MYCQZ16,Stipcevic2015,Kollmitzer2020}. The random symbols define the polarization bases for encoding the information into photons. Alice needs to store Gb/s raw symbols she used until Bob responds which photons he received and which bases he used for photon measurements. The greater the distance, the greater the round-trip delay, which affects memory capacity. The delay also affects Bob, who is waiting for information from Alice on which bases have been chosen correctly and which measurements can be considered correct. Although post-processing occurs through a public channel, its organization can set constraints on quantum channel performance. When post-processing modules are connected in series, the QKD system cannot process new photons until previously received photons are processed~\cite{Mehic2017a}.

The QKD process stands out from other key distribution techniques due to the ability to assess and detect eavesdropping. In this process, the value of the estimated quantum probability of error (QBER) is essential. If the estimated value is greater than the typical threshold value of the optical channel, it is concluded that unforeseen activities that may result from eavesdropping have occurred~\cite{Scarani2008,Gisin2002}. If the measured QBER value is less than the limit value, it is concluded that the connection is secure, and the error reconciliation step can be accessed. Additionally, the estimated QBER value dictates the settings of other post-processing modules.

Error reconciliation tasks are coarse-grain computations since they require extensive computation resources. They are efficient in a parallel processing environment which can provide linear speedup. Although parallel processing involving multiple devices is a relatively easy solution for a research environment, it is cumbersome for practical deployment. Well-known error key estimation~\cite{brassard1994} and reconciliation~\cite{MartinezMateo2013} approaches require significant computational resources and considerable processing time, which reduce the overall secret key generation rate~\cite{Mehic2017a,Tomamichel2012}.

This paper presents the SarDub19 protocol that combines error estimation and reconciliation steps and significantly improves communication and computational efficiency.

%An essential benefit of the SarDub19 is the asymmetric processing which makes it suitable for use in satellite link environments.
 
%Despite advancement that are based in the development of optical/quantum equipment, QKD heavily relies on classical sub-protocols for post-processing. 

%

%When the measured QBER is less than the threshold value, no eavesdropping in the channel is considered, and the error reconciliation step is performed to correct errors~\cite{Scarani2008,Gisin2002}. However, 

%Experiment 2 determined an upper bound for the correcting power (effectiveness) of each protocol. For this reason, it is meaningless to examine a situation where the error rate was underestimated, since the protocol would surely fail to correct all errors. However, if the error rate were overestimated, it is likely that the protocol would leak excess information to any eavesdroppers, and quantifying the amount of excess will provide a deeper understanding of the importance of error estimation and allow the development of an error rate estimation bound.

%%%%%%%%%%%%%%%%%%%%%%%%%%%%%%%%%%%%%%%%%%%%%%%% RESULTS %%%%%%%%%%%%%%%%%%%%%%%%%%%%%%%%%%%%%%%%%%%%%%%%

\section{Results}

Popular approaches for determining QBER are based on a public comparison of a part of the raw key assuming a uniform error distribution~\cite{Tomamichel2012,Dusek2006}. Also, error reconciliation solutions take up significant communication and computational resources~\cite{Bennett1992d,brassard1994,Elkouss2013b,Elkouss2009,Buttler2003,Kollmitzer2010} and lead to notable shortening of the key~\cite{Calver2011a}. Thus, the main goal of SarDub19 is to exploit QKD system features for the realization of error key estimation and reconciliation tasks with minimal computing and communication resources.

\subsection{Communication Efficiency}

Unlike Cascade, SarDub19 detects errors in only a few iterations. As the length of the sifted key $n$ and the number of errors increases, the number of iterations increases (Table~\ref{tab:comparison_iterations}). Each iteration indicates the exchange of two messages and introduces an additional delay. However, the overall number of exchanged messages and, therefore, the total delay is significantly less than Cascade and Winnow. 

\begin{table}
\caption{The average number of needed SarDub19 iterations to detect all errors vs. the length of the sifted key $n$.}
\label{tab:comparison_iterations}
\begin{tabular}{lccccc}
\hline
eff/n  & 1       & 2       & 3       & 4       & 5      \\ \hline
1000   & 14.08\% & 80.16\% & 5.7\%   & 0.06\%  & -      \\
10000  & 5.34\%  & 34.15\% & 55.11\% & 5.4\%   & -      \\
100000 & -       & 16.74\% & 55.29\% & 27.93\% & 0.04\% \\ \hline
\end{tabular}
\end{table}

\subsection{Computation Complexity}

SarDub19 offers significant reductions in terms of processing time compared to LDPC (Fig.~\ref{fig:all_time}). In addition, SarDub19 requires several randomly shuffled arrays of maximum length $n$, which is significantly less than the memory capacity needed to store LDPC generator and parity matrices. Considering the communication complexity, the size of the reconciled key, and the number of messages exchanged in the view of throughput, as shown in Fig.~\ref{fig:all_throughput}, SarDub19 outperforms existing solutions and significantly speeds up post-processing operations.

\subsection{Security}

The SarDub19 leverages the randomness of QKD keys to serve as seeds for random permutations. Assuming that eavesdropper Eve has no information about the seeds used to form random permutations, she can only obtain information about the parity-check values of randomly permuted blocks. Thus, she would gain no knowledge of the sequence order of bits in the sifted key before applying random permutations. However, to mitigate the information leakage, it would be necessary to discard one bit for each parity value exchanged~\cite{Bennett1992d} or apply privacy enhancement procedure~\cite{brassard1994}.

SarDub19 simplifies post-processing by combining error estimation and reconciliation phases. This avoids key shortening to estimate QBER values and speeds up the operation of post-processing software. Alice and Bob generally estimate error rates by selecting a set of random bits and comparing them over the public channel. These uncovered bits are dismissed since publicly disclosed information is available to Eve. The error rate is driven by noise in the channel and possible eavesdropper interference. %If the estimated error rate is higher than the tolerable error rate,  it is assumed that communication is affected by an unexpected impact such as eavesdropping and the key distribution process is terminated. 
Additionally, the estimated QBER value defines the error reconciliation technique settings such as block or matrix sizes~\cite{Calver2011a}. In SarDub19, the estimation phase is merged with the error reconciliation. Combining these two phases reduces the number of messages exchanged and the number of leaked and discarded bits.

\subsubsection{Box 1 | Protocol Definition} 	 	

\textbf{Preparation} Alice and Bob will form $m$ identical random permutations over the sifted keys of length $n$. The random permutations can be implemented using a pseudo-random number generator relying on pre-established symmetrical QKD keys as seeds. Then, the sifted key is divided into blocks containing four bits. For each permutation $j=1,2..m$ and every block $k_j=1,2,3..\frac{n}{4}$, Alice and Bob calculate parity-check values $p_{A,k_j}$ and $p_{B,k_j}$. 

%SarDub19 uses small poriton of pre-established symetrical QKD key that have the random origin to define random $m$ permutations of the sifted key of length $n$~\cite{brassard1994}. 

\textbf{Step A - First Iteration} Alice sends $p_{A,k_1}$ to Bob who compares them with his $p_{B,k_1}$ values and forms a lists of blocks with matching and mismatching parity-check values. Blocks with mismatching parity-check values contain an odd number of error bits and can be used to estimate QBER value (see Box 2). Bob informs Alice about mismatching  blocks by sending block's identifier $k_1$ and the hash value calculated over the list of bits located in matching even blocks. Alice temporarily discards mismatching blocks and calculates hash value over the list of bits in matching blocks. If hash values are identical, all errors are detected and located in the mismatching blocks. For small QBER values such as 1\%, SarDub19 identifies all errors located in mismatching blocks in the first iteration. However, additional iterations are needed to detect errors hidden in even blocks for keys with larger QBER. 

\textbf{Step B - Next Iteration} Alice applies the next $j$ of $m$ random permutations but now over the truncated sifted key without previously marked mismatching error blocks. The list of newly calculated parity check values $p_{A,k_{j>1}}$ is sent to Bob. He applies the same random permutation $j$ to his truncated sifted key and compares his sub-block parity values $p_{B,k_{j>1}}$ with the received values forming a new list of matching and mismatching blocks. Bob uses those bits that form the latest mismatching blocks and identify the matching blocks from the first iteration in which these bits were located. For each mismatching block of the last iteration, Bob discards the four matching blocks from the first iteration in which the errors were masked. The sifted key is further shortened by omitting identified blocks, and a new hash value is calculated. The identifiers $k$ of detected blocks with masked errors are provided to Alice along with the hash value. 
Alice will discard the selected blocks and calculate the hash value over the truncated sifted key. If hash values match, the protocol terminates. Otherwise, it indicates the presence of additional masked errors that need to be detected by applying the next random permutation and going back to step B.

\subsubsection{Box 2 | QBER Estimation} 	 

\textbf{Theorem 1}. The relationship between the expected value of the number of blocks with mismatching parity $\bar{m}$, the number of error bits $q$, and the number of bits in the key $n$ is given with

\begin{equation}
\bar{m}=\frac{q(n-q)(n^2-2nq-3n+2q^2+4)}{(n-3)(n-2)(n-1)}
\end{equation}

\textit{Proof}: Because of the fixed block length of 4, blocks with mismatching parity fall into two disjunct categories: (1) blocks with exactly one error and (2) blocks with exactly three errors. We calculate the estimated value of the number of blocks in each category. 

Let us observe a process in which $q$ error bits are randomly distributed over $n$ positions grouped into blocks of four. A block will contain exactly one error if and only if one error bit lands in one of its four positions and all other error bits land elsewhere. The probability for this happening is:

\begin{equation}
p_1 = \left( \prod_{a=n-q+1}^{u-1} \left( 1-\frac{3}{a} \right)\right)  \left( \frac{4}{u} \right)  \left( \prod_{b=u+1}^{n} \left( 1-\frac{4}{b} \right) \right),
\end{equation}

because we observe three stages in this scenario (written from right to left): first errors missing the four positions of the observed block, an error landing in one of the four positions, and the remaining errors missing the remaining three positions in the observed block. Probability of an error landing in a block containing $k$ errors $(k=0,...,4)$ is $(4-k)/\tilde{n}$, where $\tilde{n}$ is the number of overall vacant positions, and consequently, the probability of missing the block is $1-(4-k)/\tilde{n}$. We start with $\tilde{n}=n$ initially for the first error bit (rightmost product) and end with $\tilde{n}=n-q+1$ for the last error bit (leftmost product). The products are telescopic, so the expression reduces to

\begin{equation}
p_1=\frac{4(n-q-2)(n-q-1)(n-q)}{(n-3)(n-2)(n-1)n}
\end{equation}

As expected, the probability is independent of the particular value of $u$. Probability for all blocks is the sum of $p_1$ for all possible $u$ (i.e. all $q$ possible values of $\tilde{n}$, from $n-q+1$ to $n$, which means just multiplication $qp_1$). To obtain the expected value of the number of blocks with single errors, we multiply this probability with the number of blocks $n/4$. Thus, we obtain

\begin{equation}
P_1(n,q)=\frac{q(n-q-2)(n-q-1)(n-q)}{(n-3)(n-2)(n-1)}
\end{equation}

A block will contain exactly three errors if and only if three error bits land in three of its four positions, and all other error bits land elsewhere. The probability of this happening is

\begin{equation}
    p_3=\left( \prod_{a=n-q+1}^{u-1} (1-\frac{1}{a}) \right) \left( \frac{2}{u} \right) \left( \prod_{b=u+1}^{v-1} \left( 1-\frac{2}{b} \right) \right) \left( \frac{3}{v} \right) \left( \prod_{c=v+1}^{w-1} \left( 1-\frac{3}{c} \right) \right)\left( \frac{4}{w} \right)\left( \prod_{d=w+1}^{n} \left( 1-\frac{4}{d} \right) \right)
\end{equation}

with the reasoning same as the above. Again, the products are telescopic, so we obtain

\begin{equation}
p_3= \frac{24(n-q)}{(n-3)(n-2)(n-1)n}
\end{equation}

The probabilities are independent of the choices for $u,v,w$, and the total probability will be obtained by multiplying $p_3$ with the number of possible options, which is $\binom{q}{3}$. Once again, to obtain the expected value of the number of blocks with three errors, we multiply this probability with the number of blocks $n/4$. This yields

\begin{equation}
P_3 = \frac{q(q-1)(q-2)(n-q)}{(n-3)(n-2)(n-1)}
\end{equation}

The total expected value of blocks with mismatched parity is then the sum of the two expected values calculated above, i.e.

\begin{equation}
m=P_1+P_3=\frac{q(n-q)(n2-2nq-3n + 2q 2 + 4)}{(n-3)(n-2)(n-1)}
\end{equation}

\flushright
$\blacksquare$
\flushleft

\textit{Remark 2}. As seen above, $\bar{m}$ is a quartic polynomial in $q$ for $n=const$ and, as such, can be exactly solved on its whole domain. However, we can be approximate the part of the function relevant to our work and solve a more straightforward equation. This particular function for key length of 1024 is given by:
%Fig. 1(b) shows a third-degree polynomial approximation for $q(\bar{m})$ in case $n=1024$ on the first $20\%$ of values of $q$ (i.e. for QBER up to 0.2).

\begin{equation}
q = c_1\bar{m}^3+c_2\bar{m}^2+c_3\bar{m}+c_4
\end{equation}

where $c_1=0.00010813$, $c_2=-00778112$, $c_3=1.351$, $c_4=2.45078125$. For other key lengths we can perform curve fitting to obtain the approximations or use this one with the following modification.

\begin{equation}
q = c_1 \left( \frac{1024}{n} \right)^2 \bar{m}^3 + c_2 \left( \frac{1024}{n} \right) \bar{m}^2 + c_3\bar{m} + c_4 \left( \frac{n}{1024} \right)
\end{equation}

Our method can estimate the number of errors from as few as three samples within 5\% of the correct value, where the estimation error accounts for both the approximation error of curve fitting for more straightforward computation and the difference between the average of samples and actual expected value of the random variable.

%%%%%%%%%%%%%%%%%%%%%%%%%%%%%%%%%%%%%%%%%%%%%%%% DISCUSSION %%%%%%%%%%%%%%%%%%%%%%%%%%%%%%%%%%%%%%%%%%%%%%%%

\begin{figure}
	\includegraphics[width=1\textwidth]{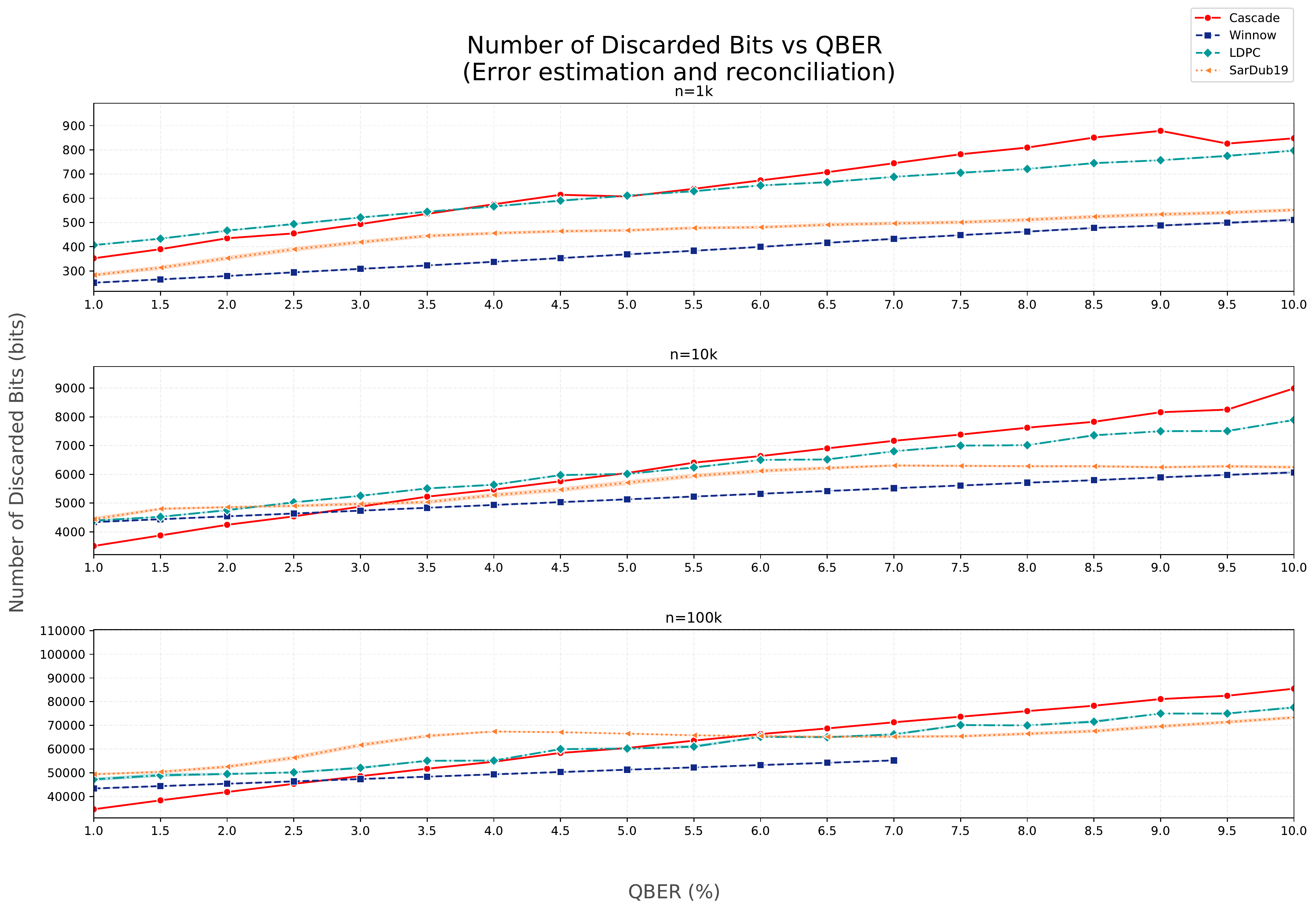}
	\caption{The number of information bits exposed and discarded for different values of QBER. The number of discarded bits for LDPC is 25\% of the sifted key used for error estimation and the length of the syndrome exchanged; for Cascade it is  25\% of the sifted key used for error estimation and the number of exchanged parity values; for Winnow it is the number of exchanged parity values and the length of syndromes exchanged; for SarDub19 it is the number of exchanged parity values and the number of discarded bits.\label{fig:all_bits_leaked}}
	
\end{figure}

\begin{figure}
	\includegraphics[width=1\textwidth]{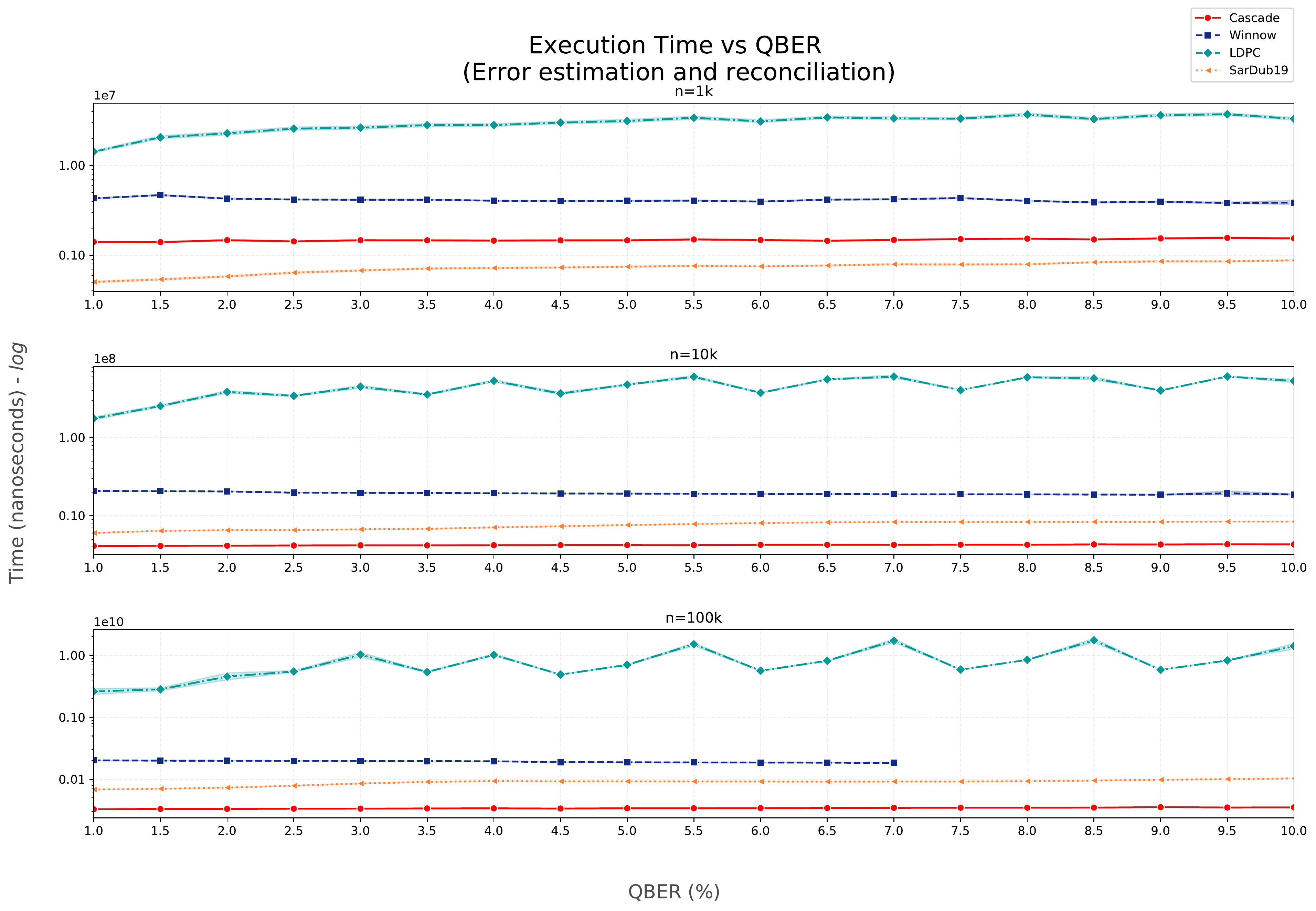}
	\caption{Execution time for different values of QBER.\label{fig:all_time}}
\end{figure}

\begin{figure}
	\centering
	\includegraphics[width=1\textwidth]{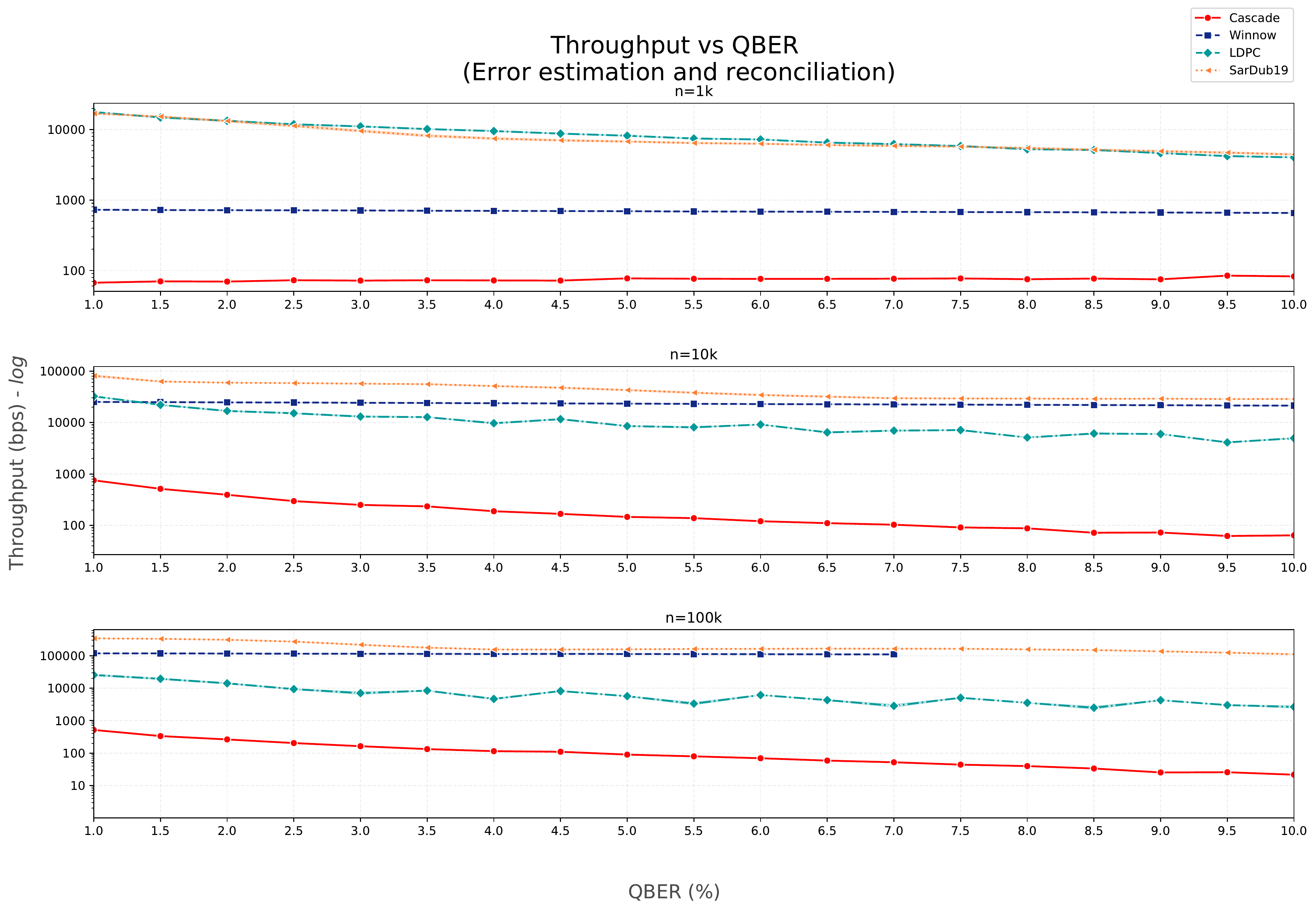}
	\caption{Overall protocol throughput depends on the number of discarded bits, the execution time of the protocol, and the number of messages exchanged.\label{fig:all_throughput}}
	
\end{figure}

\section{Discussion}

One may argue whether performing error estimation in each protocol execution is necessary. While this may improve protocol performance, skipping the error estimation compromises the security of the key distribution process as participants cannot detect anomalies that may arise from eavesdropping. 
Cascade and LDPC require a precise value of the estimated QBER  to adjust operating parameters. The estimation phase is usually independent and is calculated by publicly analyzing (and discarding) 25\% of the sifted key~\cite{Calver2011a}. Winnow and SarDub19 integrate estimation and reconciliation phases, noting that Winnow's accuracy is often insufficient, which leads to misconfiguration settings and the inability to correct errors at higher QBER values and longer keys~\cite{Buttler2003}. The accuracy of SarDub19 error estimation is significant (Pearson correlation coefficient r = 0.987) which enables reliable and timely detection of eavesdropping activities.

SarDub19 uses hash functions to verify the information exchanged. Given that SarDub19 implements part of the privacy amplification phase (through discarding mismatching blocks) with error estimation and reconciliation phases, the exchange of SarDub19 hash values corresponds to the authentication phase of standard QKD post-processing stacks~\cite{bennett1984quantum}. The hash values calculated for each iteration can be based on post-quantum cryptography, while the final hash value can be calculated using the $\epsilon-ASU_{2}$ family of hash functions such as Wegman-Carter~\cite{Cederlof2008,Portmann2014}.

In addition to the initial version that relies on recursively analyzing parity values from previous iterations, SarDub19 can be implemented to instantly discard those blocks for which misparity values are identified. Such an approach reduces the number of discarded blocks per iteration but introduces additional iterations, and more parity check values are to be discarded.

%objasniti zašto se kod Winnow-a dešava mala efikasnost (ne mogućnosti završetka protokola)
%poređenja SARDUB19 sa i bez odbacivanja u iteracijama
%Poređenja sa sigurnosnog aspekta, kakav hash je neophodan za provjeru rada?

%%%%%%%%%%%%%%%%%%%%%%%%%%%%%%%%%%%%%%%%%%%%%%%% METHODS %%%%%%%%%%%%%%%%%%%%%%%%%%%%%%%%%%%%%%%%%%%%%%%%

\section{Methods}

%Compare winnow number of discarded, leaked bits with others
%Compare number of successful reconciliation processes with others
%Replot time and throughput (throughput over reconciled key: sifted-discarded)

\subsection{Correctness}

The performances of SarDub19 protocol were compared with Cascade, Winnow, and LDPC techniques evaluating the execution time, the number of exposed/discarded bits, and the overall throughput. %Implementation of each protocol in C++ was developed using a common structure and data types.
We performed executions for different lengths of sifted key ($n=1000, 10000, 100000$ bits) and QBER values ($qber=1, 1.5, 2, 2.5,...10 \%$). All experiments were executed using 1000 random seeds per single QBER value. 

%229780*4(sardub,cascade,winnow,ldpc)*2(1k,10k) + 23000*4(sardub,cascade,winnow,ldpc)=1930240
 
%Here is a description of a specific method used.  Note that the
%subsection heading ends with a full stop (period) and that the
%command is \verb|\subsection{}| not \verb|\subsection*{}|.

The C++ source code of Cascade, Winnow, and LDPC is taken from~\cite{Johnson2012}. Each message exchanged between Alice and Bob was counted as a 20 ms delay to calculate processing time. The total number of messages exchanged is minimized in implementation to compare protocols evenly. For example, instead of sending each parity check value separately, in initial iteration, all parity check values can be exchanged in one message. The summarized delay is used in calculating the protocol throughput shown in Figure~\ref{fig:all_throughput}. However, the execution time of the protocol is shown in Figure~\ref{fig:all_time}.

\paragraph{Cascade}

The fluctuations of reconciled key length are based on the distribution of errors. For Cascade, the starting block size is an essential parameter. If the starting block is too large, the algorithm may not be able to detect all errors. However, when the starting block is too small, it may result in an unnecessarily large number of exchanged messages leading to increased communication traffic, longer processing times, and more discarded bits in the privacy amplification phase. %Therefore, the suitable starting block size is the largest size that still results in all errors being corrected.
The well-known analytical expression $k=0.73/QBER_{estimated}$ was used in our simulations. The estimated QBER was passed as a simulation parameter, with the length of the discarded bits within the QBER estimation calculated as 25\% of the sifted key~\cite{Calver2011a}.

%Furthermore, in order to account for network latency in a uniform way, 20 ms are added to the runtime calculation for each message passed between Alice and Bob. For Cascade, this means one message for each block parity exchange, and two messages for each parity bit exchange (Bob must send his parity to Alice, and she must respond with match or mismatch). In order to track the number of messages exchanged, a separate, messageCount variable is maintained while the algorithm is running, and 20 x messageCount milliseconds are added to the runtime of each of the iterations.

%Also, for this experiment, network latency is not considered empirically, and the experiment is conducted on one machine. The latency of the network is assumed to be 20 ms, which is a reasonable if not optimistic assumption given the current state of technology. Furthermore, the number of messages exchanged between Alice and Bob are highly dependent on the implementation. For instance, one implementation may have Alice and Bob exchange all their block parities in one message while another would separate the parities into individual messages. The method selected here takes a conservative measure, and assumes the minimum number of messages is exchanged.

\paragraph{Winnow} 

The Winnow source-code was updated to include the error estimation step. The number of odd and even parity values can be used to estimate the QBER~\cite{Buttler2003}. However, one iteration is often not enough to collect a sufficient number of samples for precise estimation and reconciliation. %Thus, several iterations are performed where the number of iterations depends on the settings of the Winnow protocol. 
The Winnow QBER estimation source code is taken from~\cite{Kevin2010}.

\paragraph{LDPC} 

The LDPC codes were generated using the progressive growth algorithm~\cite{Johnson2015}. %It may be possible to generate codes with better performance characteristics; however, the purpose here was to examine the general performance of an algorithm based on LDPC codes versus other protocols and not on achieving the maximum performance possible. 
The estimated QBER was passed as a simulation parameter, with the amount of the discarded bits within the QBER estimation phase calculated as 25\% of the sifted key~\cite{Calver2011a}.

\subsection{Data availability}

The datasets generated and analysed during the current study are available from the corresponding author on reasonable request.

%\begin{methods}
%Put methods in here.  If you are going to subsection it, use
%\verb|\subsection| commands.  Methods section should be less than
%800 words and if it is less than 200 words, it can be incorporated into the main text.
%\end{methods}

\subsection{Competing Interests}

%M.M. and H.S. are inventors of Patent 2208318.2.
Authors are named as inventors on patent relating to SarDub19 protocol.

\section{Author contribution}

%Both authors developed the main ideas. M.M. formulated the technical claims, provided numerical simulations, and wrote the manuscript. H.S. provided analytical formulations and contributed to the technical derivations and write-up.

Both authors developed the main ideas. M.Mehic formulated the technical claims, provided numerical simulations, and wrote the manuscript. H.Siljak provided analytical formulations and contributed to the technical derivations and write-up.

%\section{Figure legends}

\bibliographystyle{unsrtnat}
%\bibliography{references}  

\begin{thebibliography}{0}
\providecommand{\natexlab}[1]{#1}
\providecommand{\url}[1]{\texttt{#1}}
\expandafter\ifx\csname urlstyle\endcsname\relax
  \providecommand{\doi}[1]{doi: #1}\else
  \providecommand{\doi}{doi: \begingroup \urlstyle{rm}\Url}\fi

\end{thebibliography}


\begin{thebibliography}{40}
\providecommand{\natexlab}[1]{#1}
\providecommand{\url}[1]{\texttt{#1}}
\expandafter\ifx\csname urlstyle\endcsname\relax
  \providecommand{\doi}[1]{doi: #1}\else
  \providecommand{\doi}{doi: \begingroup \urlstyle{rm}\Url}\fi

\bibitem[Bennett et~al.(1984)Bennett, Brassard, and Others]{bennett1984quantum}
Charles~H. Bennett, Gilles Brassard, and Others.
\newblock {Quantum cryptography: Public key distribution and coin tossing}.
\newblock In \emph{Proceedings of IEEE International Conference on Computers,
  Systems and Signal Processing}, volume 175, page~8. New York, Elsevier B.V.,
  1984.
\newblock \doi{10.1016/j.tcs.2011.08.039}.

\bibitem[Brassard et~al.(1994)Brassard, Salvail, Louis, Salvail, and
  Louis]{brassard1994}
Gilles Brassard, Louis Salvail, Salvail Louis, Louis Salvail, and Salvail
  Louis.
\newblock {Secret-key reconciliation by public discussion}.
\newblock \emph{Advances in Cryptology - EUROCRYPT93}, 765:\penalty0 410--423,
  1994.
\newblock \doi{10.1007/3-540-48285-7_35}.

\bibitem[Mehic et~al.(2020)Mehic, Niemiec, Rass, Ma, Peev, Aguado, Martin,
  Schauer, Poppe, Pacher, and Voznak]{Mehic2020b}
Miralem Mehic, Marcin Niemiec, Stefan Rass, Jiajun Ma, Momtchil Peev, Alejandro
  Aguado, Vicente Martin, Stefan Schauer, Andreas Poppe, Christoph Pacher, and
  Miroslav Voznak.
\newblock {Quantum Key Distribution}.
\newblock \emph{ACM Computing Surveys}, 53\penalty0 (5):\penalty0 1--41, oct
  2020.
\newblock ISSN 0360-0300.
\newblock \doi{10.1145/3402192}.

\bibitem[Sasaki et~al.(2011)Sasaki, Fujiwara, Ishizuka, Klaus, Wakui, Takeoka,
  and Miki]{SasakiM20111}
M.~Sasaki, M.~Fujiwara, H.~Ishizuka, W.~Klaus, K.~Wakui, M.~Takeoka, and
  S.~Miki.
\newblock {Field Test of Quantum Key Distribution in the Tokyo QKD Network}.
\newblock \emph{Optics Express}, 19\penalty0 (2011), aug 2011.

\bibitem[Peev et~al.(2009)Peev, Pacher, All{\'{e}}aume, Barreiro, Bouda,
  Boxleitner, Debuisschert, Diamanti, Dianati, Dynes, Fasel, Fossier,
  F{\"{u}}rst, Gautier, Gay, Gisin, Grangier, Happe, Hasani, Hentschel,
  H{\"{u}}bel, Humer, L{\"{a}}nger, Legr{\'{e}}, Lieger, Lodewyck,
  Lor{\"{u}}nser, L{\"{u}}tkenhaus, Marhold, Matyus, Maurhart, Monat, Nauerth,
  Page, Poppe, Querasser, Ribordy, Robyr, Salvail, Sharpe, Shields, Stucki,
  Suda, Tamas, Themel, Thew, Thoma, Treiber, Trinkler, Tualle-Brouri, Vannel,
  Walenta, Weier, Weinfurter, Wimberger, Yuan, Zbinden, and
  Zeilinger]{Peev2009}
M.~Peev, C.~Pacher, R.~All{\'{e}}aume, C.~Barreiro, J.~Bouda, W.~Boxleitner,
  T.~Debuisschert, E.~Diamanti, M.~Dianati, J.~F. Dynes, S.~Fasel, S.~Fossier,
  M.~F{\"{u}}rst, J-D Gautier, O.~Gay, N.~Gisin, P.~Grangier, A.~Happe,
  Y.~Hasani, M.~Hentschel, H.~H{\"{u}}bel, G.~Humer, T.~L{\"{a}}nger,
  M.~Legr{\'{e}}, R.~Lieger, J.~Lodewyck, T.~Lor{\"{u}}nser,
  N.~L{\"{u}}tkenhaus, A.~Marhold, T.~Matyus, O.~Maurhart, L.~Monat,
  S.~Nauerth, J-B Page, A.~Poppe, E.~Querasser, G.~Ribordy, S.~Robyr,
  L.~Salvail, A.~W. Sharpe, A.~J. Shields, D.~Stucki, M.~Suda, C.~Tamas,
  T.~Themel, R.~T. Thew, Y.~Thoma, A.~Treiber, P.~Trinkler, R.~Tualle-Brouri,
  F.~Vannel, N.~Walenta, H.~Weier, H.~Weinfurter, I.~Wimberger, Z.~L. Yuan,
  H.~Zbinden, and A.~Zeilinger.
\newblock {The SECOQC quantum key distribution network in Vienna}.
\newblock \emph{New Journal of Physics}, 11\penalty0 (7):\penalty0 075001, jul
  2009.
\newblock ISSN 1367-2630.
\newblock \doi{10.1088/1367-2630/11/7/075001}.

\bibitem[Chen et~al.(2009)Chen, Liang, Liu, Cai, Ju, Liu, Wang, Yin, Chen,
  Chen, Peng, and Pan]{chen2009}
Teng-Yun Chen, Hao Liang, Yang Liu, Wen-Qi Cai, Lei Ju, Wei-Yue Liu, Jian Wang,
  Hao Yin, Kai Chen, Zeng-Bing Chen, Cheng-Zhi Peng, and Jian-Wei Pan.
\newblock {Field Test of a Practical Secure Communication Network With
  Decoy-State Quantum Cryptography}.
\newblock \emph{Optics Express}, 17\penalty0 (8):\penalty0 6540, apr 2009.
\newblock ISSN 1094-4087.
\newblock \doi{10.1364/OE.17.006540}.

\bibitem[Wang et~al.(2014)Wang, Chen, Yin, Li, He, Li, Zhou, Song, Li, Wang,
  Chen, Han, Huang, Guo, Hao, Li, Zhang, Liu, Liang, Miao, Wu, Guo, and
  Han]{Wang2014}
Shuang Wang, Wei Chen, Zhen-Qiang Yin, Hong-Wei Li, De-Yong He, Yu-Hu Li, Zheng
  Zhou, Xiao-Tian Song, Fang-Yi Li, Dong Wang, Hua Chen, Yun-Guang Han,
  Jing-Zheng Huang, Jun-Fu Guo, Peng-Lei Hao, Mo~Li, Chun-Mei Zhang, Dong Liu,
  Wen-Ye Liang, Chun-Hua Miao, Ping Wu, Guang-Can Guo, and Zheng-Fu Han.
\newblock {Field and long-term demonstration of a wide area quantum key
  distribution network}.
\newblock \emph{Optics Express}, 22\penalty0 (18):\penalty0 21739, sep 2014.
\newblock ISSN 1094-4087.
\newblock \doi{10.1364/OE.22.021739}.

\bibitem[Elliott et~al.(2005)Elliott, Colvin, Pearson, Pikalo, Schlafer, and
  Yeh]{Elliott2005}
Chip Elliott, Alexander Colvin, David Pearson, Oleksiy Pikalo, John Schlafer,
  and Henry Yeh.
\newblock {Current status of the DARPA quantum network (Invited Paper)}.
\newblock In Eric~J Donkor, Andrew~R Pirich, and Howard~E Brandt, editors,
  \emph{Proc. SPIE 5815, Quantum Information and Computation III,}, volume
  5815, pages 138--149, may 2005.
\newblock \doi{10.1117/12.606489}.

\bibitem[Hosseinidehaj et~al.(2018)Hosseinidehaj, Babar, Malaney, Ng, and
  Hanzo]{Hosseinidehaj2018}
Nedasadat Hosseinidehaj, Zunaira Babar, Robert Malaney, Soon~Xin Ng, and Lajos
  Hanzo.
\newblock {Satellite-Based Continuous-Variable Quantum Communications:
  State-of-the-Art and a Predictive Outlook}.
\newblock \emph{IEEE Communications Surveys and Tutorials}, PP:\penalty0 1,
  2018.
\newblock ISSN 1553877X.
\newblock \doi{10.1109/COMST.2018.2864557}.

\bibitem[Lucamarini et~al.(2018)Lucamarini, Yuan, Dynes, and
  Shields]{Lucamarini2018a}
M.~Lucamarini, Z.~L. Yuan, J.~F. Dynes, and A.~J. Shields.
\newblock {Overcoming the rate–distance limit of quantum key distribution
  without quantum repeaters}.
\newblock \emph{Nature}, 557\penalty0 (7705):\penalty0 400--403, may 2018.
\newblock ISSN 0028-0836.
\newblock \doi{10.1038/s41586-018-0066-6}.

\bibitem[Wehner et~al.(2018)Wehner, Elkouss, and Hanson]{Wehner2018a}
Stephanie Wehner, David Elkouss, and Ronald Hanson.
\newblock {Quantum internet: A vision for the road ahead}.
\newblock \emph{Science}, 362\penalty0 (6412), oct 2018.
\newblock ISSN 0036-8075.
\newblock \doi{10.1126/science.aam9288}.

\bibitem[Korzh et~al.(2015)Korzh, Lim, Houlmann, Gisin, Li, Nolan, Sanguinetti,
  Thew, and Zbinden]{Korzh2015}
Boris Korzh, Charles Ci~Wen Lim, Raphael Houlmann, Nicolas Gisin, Ming~Jun Li,
  Daniel Nolan, Bruno Sanguinetti, Rob Thew, and Hugo Zbinden.
\newblock {Provably secure and practical quantum key distribution over 307 km
  of optical fibre}.
\newblock \emph{Nature Photonics}, 9\penalty0 (3):\penalty0 163--168, feb 2015.
\newblock ISSN 1749-4885.
\newblock \doi{10.1038/nphoton.2014.327}.

\bibitem[Shor and Preskill(2000)]{Shor2000}
Peter W.~Pw Shor and John Preskill.
\newblock {Simple proof of security of the BB84 quantum key distribution
  protocol}.
\newblock \emph{Physical review letters}, 85\penalty0 (2):\penalty0 441--444,
  jul 2000.
\newblock ISSN 1079-7114.
\newblock \doi{10.1103/PhysRevLett.85.441}.

\bibitem[Mayers(2001)]{Mayers2001}
Dominic Mayers.
\newblock {Unconditional Security in Quantum Cryptography}.
\newblock \emph{Journal of the ACM}, 48\penalty0 (3):\penalty0 351--406, 2001.
\newblock ISSN 00045411.
\newblock \doi{10.1145/382780.382781}.

\bibitem[Renner et~al.(2005)Renner, Gisin, and Kraus]{Renner2005a}
Renato Renner, Nicolas Gisin, and Barbara Kraus.
\newblock {Information-theoretic security proof for quantum-key-distribution
  protocols}.
\newblock \emph{Physical Review A - Atomic, Molecular, and Optical Physics},
  2005.
\newblock ISSN 10502947.
\newblock \doi{10.1103/PhysRevA.72.012332}.

\bibitem[Gisin et~al.(2002{\natexlab{a}})Gisin, Ribordy, Tittel, and
  Zbinden]{gisin2002quantum}
Nicolas Gisin, Gr{\'e}goire Ribordy, Wolfgang Tittel, and Hugo Zbinden.
\newblock Quantum cryptography.
\newblock \emph{Reviews of modern physics}, 74\penalty0 (1):\penalty0 145,
  2002{\natexlab{a}}.

\bibitem[Bedington et~al.(2017)Bedington, Arrazola, and Ling]{Bedington2017a}
Robert Bedington, Juan~Miguel Arrazola, and Alexander Ling.
\newblock {Progress in satellite quantum key distribution}.
\newblock \emph{npj Quantum Information}, 3\penalty0 (1):\penalty0 30, dec
  2017.
\newblock ISSN 2056-6387.
\newblock \doi{10.1038/s41534-017-0031-5}.

\bibitem[Liao et~al.(2017)Liao, Cai, Liu, Zhang, Li, Ren, Yin, Shen, Cao, Li,
  Li, Chen, Sun, Jia, Wu, Jiang, Wang, Huang, Wang, Zhou, Deng, Xi, Ma, Hu,
  Zhang, Chen, Liu, Wang, Zhu, Lu, Shu, Peng, Wang, and Pan]{Liao2017a}
Sheng-Kai Liao, Wen-Qi Cai, Wei-Yue Liu, Liang Zhang, Yang Li, Ji-Gang Ren,
  Juan Yin, Qi~Shen, Yuan Cao, Zheng-Ping Li, Feng-Zhi Li, Xia-Wei Chen, Li-Hua
  Sun, Jian-Jun Jia, Jin-Cai Wu, Xiao-Jun Jiang, Jian-Feng Wang, Yong-Mei
  Huang, Qiang Wang, Yi-Lin Zhou, Lei Deng, Tao Xi, Lu~Ma, Tai Hu, Qiang Zhang,
  Yu-Ao Chen, Nai-Le Liu, Xiang-Bin Wang, Zhen-Cai Zhu, Chao-Yang Lu, Rong Shu,
  Cheng-Zhi Peng, Jian-Yu Wang, and Jian-Wei Pan.
\newblock {Satellite-to-ground quantum key distribution}.
\newblock \emph{Nature}, 549\penalty0 (7670):\penalty0 43--47, sep 2017.
\newblock ISSN 0028-0836.
\newblock \doi{10.1038/nature23655}.

\bibitem[Yin et~al.(2017)Yin, Cao, Li, Liao, Zhang, Ren, Al., Liu, {Bo Li}, Li,
  Lu, Gong, Xu, Li, Li, Yin, Jiang, Li, Jia, {Ge Ren}, Zhou, Zhang, Wang,
  Chang, Zhu, Liu, Chen, Lu, Shu, Peng, Wang, and Pan]{Yin2017}
Juan Yin, Yuan Cao, Yu-Huai Li, Sheng-Kai Liao, Liang Zhang, Ji-Gang Ren,
  Wen-Qi~Cai Al., Wei-Yue Liu, Hui~Dai {Bo Li}, Guang-Bing Li, Qi-Ming Lu,
  Yun-Hong Gong, Yu~Xu, Shuang-Lin Li, Feng-Zhi Li, Ya-Yun Yin, Zi-Qing Jiang,
  Ming Li, Jian-Jun Jia, Dong~He {Ge Ren}, Yi-Lin Zhou, Xiao-Xiang Zhang,
  Na~Wang, Xiang Chang, Zhen-Cai Zhu, Nai-Le Liu, Yu-Ao Chen, Chao-Yang Lu,
  Rong Shu, Cheng-Zhi Peng, Jian-Yu Wang, and Jian-Wei Pan.
\newblock {Satellite-based entanglement distribution over 1200 kilometers}.
\newblock \emph{Science}, 356\penalty0 (6343):\penalty0 1140--1144, 2017.
\newblock \doi{10.1126/science.aan3211}.

\bibitem[Vallone et~al.(2015)Vallone, Bacco, Dequal, Gaiarin, Luceri, Bianco,
  and Villoresi]{Vallone2014}
Giuseppe Vallone, Davide Bacco, Daniele Dequal, Simone Gaiarin, Vincenza
  Luceri, Giuseppe Bianco, and Paolo Villoresi.
\newblock {Experimental Satellite Quantum Communications}.
\newblock \emph{Physical Review Letters}, 115\penalty0 (4):\penalty0 040502,
  jul 2015.
\newblock ISSN 0031-9007.
\newblock \doi{10.1103/PhysRevLett.115.040502}.

\bibitem[Ma et~al.(2016)Ma, Yuan, Cao, B., and Zhang]{MYCQZ16}
X~Ma, X~Yuan, Z~Cao, Qi~B., and Z~Zhang.
\newblock {Quantum random number generation}.
\newblock \emph{Nature}, 2016.

\bibitem[Stip{\v{c}}evi{\'{c}} and Bowers(2015)]{Stipcevic2015}
M.~Stip{\v{c}}evi{\'{c}} and J.~E. Bowers.
\newblock {Spatio-temporal optical random number generator}.
\newblock \emph{Optics Express}, 23\penalty0 (9):\penalty0 11619, 2015.
\newblock \doi{10.1364/oe.23.011619}.

\bibitem[Kollmitzer et~al.(2020)Kollmitzer, Schauer, Rass, and
  Rainer]{Kollmitzer2020}
Christian Kollmitzer, Stefan Schauer, Stefan Rass, and Benjamin Rainer.
\newblock \emph{{Quantum Random Number Generation}}, volume~2 of \emph{Quantum
  Science and Technology}.
\newblock Springer International Publishing, Cham, 2020.
\newblock ISBN 978-3-319-72594-9.
\newblock \doi{10.1007/978-3-319-72596-3}.

\bibitem[Mehic et~al.(2017)Mehic, Maurhart, Rass, Komosny, Rezac, and
  Voznak]{Mehic2017a}
Miralem Mehic, Oliver Maurhart, Stefan Rass, Dan Komosny, Filip Rezac, and
  Miroslav Voznak.
\newblock {Analysis of the Public Channel of Quantum Key Distribution Link}.
\newblock \emph{IEEE Journal of Quantum Electronics}, 53\penalty0 (5):\penalty0
  1--8, oct 2017.
\newblock ISSN 0018-9197.
\newblock \doi{10.1109/JQE.2017.2740426}.

\bibitem[Scarani et~al.(2009)Scarani, Bechmann-Pasquinucci, Cerf, Du{\v{s}}ek,
  L{\"{u}}tkenhaus, and Peev]{Scarani2008}
Valerio Scarani, Helle Bechmann-Pasquinucci, Nicolas~J Cerf, Miloslav
  Du{\v{s}}ek, Norbert L{\"{u}}tkenhaus, and Momtchil Peev.
\newblock {The security of practical quantum key distribution}.
\newblock \emph{Reviews of Modern Physics}, 81\penalty0 (3):\penalty0
  1301--1350, sep 2009.
\newblock ISSN 0034-6861.
\newblock \doi{10.1103/RevModPhys.81.1301}.

\bibitem[Gisin et~al.(2002{\natexlab{b}})Gisin, Ribordy, Tittel, and
  Zbinden]{Gisin2002}
Nicolas Gisin, Gregoire Ribordy, Wolfgang Tittel, and Hugo Zbinden.
\newblock {Quantum Cryptography}.
\newblock \emph{Reviews of Modern Physics}, 74\penalty0 (1):\penalty0 145--195,
  jan 2002{\natexlab{b}}.
\newblock ISSN 00346861.
\newblock \doi{10.1103/RevModPhys.74.145}.

\bibitem[Martinez-Mateo et~al.(2013)Martinez-Mateo, Elkouss, and
  Martin]{MartinezMateo2013}
Jesus Martinez-Mateo, David Elkouss, and Vicente Martin.
\newblock {Key reconciliation for high performance quantum key distribution}.
\newblock \emph{Scientific Reports}, 3:\penalty0 3--8, 2013.
\newblock ISSN 20452322.
\newblock \doi{10.1038/srep01576}.

\bibitem[Tomamichel et~al.(2012)Tomamichel, Lim, Gisin, and
  Renner]{Tomamichel2012}
Marco Tomamichel, Charles Ci~Wen Lim, Nicolas Gisin, and Renato Renner.
\newblock {Tight finite-key analysis for quantum cryptography.}
\newblock \emph{Nature communications}, 3\penalty0 (May 2011):\penalty0 634,
  jan 2012.
\newblock ISSN 2041-1723.
\newblock \doi{10.1038/ncomms1631}.

\bibitem[Dusek et~al.()Dusek, Lutkenhaus, and Hendrych]{Dusek2006}
Miloslav Dusek, Norbert Lutkenhaus, and Martin Hendrych.
\newblock In \emph{Progress in Optics}.
\newblock \doi{10.1016/S0079-6638(06)49005-3}.

\bibitem[Bennett et~al.(1992)Bennett, Bessette, Brassard, Salvail, and
  Smolin]{Bennett1992d}
Charles~H. Bennett, Fran{\c{c}}ois Bessette, Gilles Brassard, Louis Salvail,
  and John Smolin.
\newblock {Experimental quantum cryptography}.
\newblock \emph{Journal of Cryptology}, 1992.
\newblock ISSN 09332790.
\newblock \doi{10.1007/BF00191318}.

\bibitem[Elkouss et~al.(2013)Elkouss, Martinez-Mateo, and Martin]{Elkouss2013b}
David Elkouss, Jesus Martinez-Mateo, and Vicente Martin.
\newblock {Analysis of a rate-adaptive reconciliation protocol and the effect
  of leakage on the secret key rate}.
\newblock \emph{Physical Review A - Atomic, Molecular, and Optical Physics},
  87\penalty0 (4):\penalty0 1--7, 2013.
\newblock ISSN 10502947.
\newblock \doi{10.1103/PhysRevA.87.042334}.

\bibitem[Elkouss et~al.(2009)Elkouss, Leverrier, Alleaume, and
  Boutros]{Elkouss2009}
David Elkouss, Anthony Leverrier, Romain Alleaume, and Joseph~J Boutros.
\newblock {Efficient Reconciliation Protocol for Discrete-Variable Quantum Key
  Distribution}.
\newblock In \emph{2009 IEEE International Symposium on Information Theory},
  pages 1879--1883. IEEE, jun 2009.
\newblock ISBN 978-1-4244-4312-3.
\newblock \doi{10.1109/ISIT.2009.5205475}.

\bibitem[Buttler et~al.(2003)Buttler, Lamoreaux, Torgerson, Nickel, Donahue,
  and Peterson]{Buttler2003}
William~T Buttler, Steven~K Lamoreaux, Justin~R Torgerson, G~H Nickel, C.~H.
  Donahue, and Charles~G Peterson.
\newblock {Fast, efficient error reconciliation for quantum cryptography}.
\newblock \emph{Physical Review A}, 67\penalty0 (5):\penalty0 052303, may 2003.
\newblock ISSN 1050-2947.
\newblock \doi{10.1103/PhysRevA.67.052303}.

\bibitem[Kollmitzer and Pivk(2010)]{Kollmitzer2010}
Christian Kollmitzer and Mario Pivk.
\newblock \emph{{Applied Quantum Cryptography}}, volume 797.
\newblock Springer Science \& Business Media, 2010.
\newblock ISBN 3642048293.
\newblock \doi{10.1007/978-3-642-04831-9}.
\newblock URL \url{http://books.google.com/books?id=tHhqRg6lFZkC}.

\bibitem[Calver et~al.(2011)Calver, Grimaila, and Humphries]{Calver2011a}
Timothy Calver, Michael Grimaila, and Jeffrey Humphries.
\newblock {An empirical analysis of the cascade error reconciliation protocol
  for quantum key distribution}.
\newblock \emph{ACM International Conference Proceeding Series}, pages 11--14,
  2011.
\newblock \doi{10.1145/2179298.2179363}.

\bibitem[Cederl{\"{o}}f and Larsson(2008)]{Cederlof2008}
J{\"{o}}rgen Cederl{\"{o}}f and J~a. Larsson.
\newblock {Security Aspects of the Authentication used in Quantum
  Cryptography}.
\newblock \emph{IEEE Transactions on Information Theory}, 54\penalty0
  (4):\penalty0 1735--1741, 2008.
\newblock ISSN 00189448.
\newblock \doi{10.1109/TIT.2008.917697}.

\bibitem[Portmann(2014)]{Portmann2014}
Christopher Portmann.
\newblock {Key Recycling in Authentication}.
\newblock \emph{IEEE Transactions on Information Theory}, 60\penalty0
  (7):\penalty0 4383--4396, 2014.
\newblock ISSN 00189448.
\newblock \doi{10.1109/TIT.2014.2317312}.

\bibitem[Johnson(2012)]{Johnson2012}
James~S. Johnson.
\newblock \emph{{An Analysis of Error Reconciliation Protocols for Use in
  Quantum Key Distribution}}.
\newblock PhD thesis, 2012.
\newblock URL \url{https://books.google.ba/books?id=T56HMwEACAAJ}.

\bibitem[Lustic(2010)]{Kevin2010}
Kevin Lustic.
\newblock \emph{{Performance Analysis and Optimization of the Winnow Secret Key
  Reconciliation Protocol}}.
\newblock PhD thesis, Air Force Institute of Technology, 2010.
\newblock URL \url{https://books.google.ba/books?id=k18LMwEACAAJ}.

\bibitem[Johnson et~al.(2015)Johnson, Grimaila, Humphries, and
  Baumgartner]{Johnson2015}
James~S. Johnson, Michael~R. Grimaila, Jeffrey~W. Humphries, and Gerald~B.
  Baumgartner.
\newblock {An analysis of error reconciliation protocols used in Quantum Key
  Distribution systems}.
\newblock \emph{Journal of Defense Modeling and Simulation}, 12\penalty0
  (3):\penalty0 217--227, 2015.
\newblock ISSN 1557380X.
\newblock \doi{10.1177/1548512913503418}.

\end{thebibliography}

\end{document}